\documentclass[]{raa}                  
\usepackage{graphicx,times}             

\begin{document}

   \title{The Photometric System of Tsinghua-NAOC 80-cm Telescope at
   NAOC Xinglong Observatory
}

   \volnopage{Vol.0 (200x) No.0, 000--000}      
   \setcounter{page}{1}           

   \author{Fang Huang
      \inst{1,2}
   \and Jun-Zheng Li
      \inst{2}
   \and Xiao-Feng Wang
      \inst{2}
   \and Ren-Cheng Shang
      \inst{2}
   \and Tian-Meng Zhang
      \inst{3,4}
   \and Jing-Yao Hu
      \inst{3}
   \and Yu-Lei Qiu
      \inst{3}
   \and Xiao-Jun Jiang
      \inst{3,4}
        }
   \offprints{Fang Huang}                   

   \institute{Department of Astronomy, Beijing Normal University, Beijing, 100875, China; \\
  {\it hfbnu111@gmail.com} \\
        \and
             Physics Department and Tsinghua Center for Astrophysics, Tsinghua University, Beijing, 100084, China; \\
         \and
             National Astronomical Observatories, Chinese Academy of Sciences, Beijing, 100012, China;\\
         \and
        Key Laboratory of Optical Astronomy, National Astronomical Observatories, Chinese Academy of Sciences, Beijing, 100012, China\\
          }


   \abstract{
   Tsinghua-NAOC (National Astronomical Observatories of China) Telescope
   (hereafter, TNT) is an 80-cm Cassegrain reflecting telescope located
   at Xinglong Observatory of NAOC, with main scientific goals
   of monitoring various transients in the universe such as supernovae,
   gamma-ray bursts, novae, variable stars, and active galactic nuclei. We present
   in this paper a systematic test and analysis of the photometric performance of this
   telescope. Based on the calibration observations on twelve photometric nights,
   spanning the period from year 2004 to year 2012, we derived an accurate
   transformation relationship between the instrumental $ubvri$ magnitudes
   and standard Johnson $UBV$ and Cousins $RI$ magnitudes. In particular, the
   color terms and the extinction coefficients of different passbands are
   well determined. With these data, we also obtained the limiting magnitudes and
   the photometric precision of TNT. It is worthwhile to point out that the
   sky background at Xinglong Observatory may become gradually worse over the period
   from year 2005 to year 2012 (e.g., $\sim$21.4 mag vs. $\sim$20.1 mag in the V band).
   }

   \authorrunning{Huang F., J. Z. Li, X. F. Wang, R. C. Shang,  T. M. Zhang, J. Y. Hu, Y. L. Qiu, \& X. J. Jiang,}
   \titlerunning{The Photometric System of Tsinghua-NAOC Telescope}  

   \maketitle

%
%
\section{Introduction}           
\label{sect:intro}

   Tsinghua-NAOC telescope (TNT), the first professional telescope owned by a
   university in China, is an 80-cm Cassegrain telescope made by APM-Telescopes
   \footnote{$http://www.apm-telescopes.com$} in Germany. This telescope is located at Xinglong
   Observatory of NAOC ($117\dg34'39''E, 40\dg23'40''N$, with an elevation of $\sim$830m),
   jointly operated by Tsinghua University and NAOC Chinese Academy of Sciences
   since year 2004. The main sciences conducted with  TNT in recent years
   are multi-color, photometric studies of supernovae (SNe, Wang et al. \cite{Wang08}
   \& \cite{Wang09} \& \cite{Wang12}, Zhang et al. \cite{Zhang10} \& \cite{Zhang12}),
   active galactic nuclei (AGN, Liu et al. \cite{Liu10}, Zhai et al. \cite{Zhai11} \& \cite{Zhai12}),
   gamma-ray bursts (GRBs, Xin et al. \cite{Xin10} \& \cite{Xin11}). Other
   projects of this telescope involve photometric observations of binary stars (Li et al. \cite{Li09},
   Yang et al. \cite{Yang10}, Yan et al. \cite{Yan12}, Fang et al. \cite{Fang12}) and variable stars
   (Wu et al. \cite{Wu05} \& \cite{Wu06}, and Fu et al. \cite{Fu09}). However, an overall examination of
   the performance of the CCD photometric system on TNT is still absent.

   Knowing the properties and performance of a telescope such as throughput, detection limit, and
   instrument response will be of  great assistance to the observers in preparing their observation
   proposals. These parameters allow a better estimate of the exposure time and predict
   photometric precision for individual objects. We started a program to investigate
   the characteristics of the CCD photometric system on TNT. Relevant evaluations of the photometric
   system of the BATC 60/90-cm schmidt telescope and the 85-cm telescope at Xinglong Observatory of
   NAOC are available from Yan et al. (\cite{Yan00}) and Zhou et al. (\cite{Zhou09}), which help the users
   better understand the performance of these facilities and work out a reasonable observing plan.

   This paper is organized as follows:~ in Sect.~\ref{sect:inst}, we describe
   briefly about the observation system of TNT. Then we present the test results about
   the CCD detectors in Sect.~\ref{sect:ccdtest}. The photometric calibration
   results are given in Sect.~\ref{sect:phot}. The systematic performance of TNT
   is addressed in Sect.~\ref{sect:syst}, and we make a summary in Sect.~\ref{sect:sum}.


\section{Observation system}
\label{sect:inst}

TNT is a f/10 'classical' cassegrain, equatorial reflector. This telescope has a parabolic
primary mirror with an effective diameter of 0.80 m, and a hyperbolic secondary mirror with an
effective diameter of 0.26 m. The pointing of TNT is relatively fast and accurate, with the maximal
slew speed being up to 4 degrees per second. At latitudes larger than 25 degrees, the pointing accuracy
is better than $30^{\prime\prime}$. The pointing drift without guide star tracking is less than $1^{\prime\prime}$
in 15 minutes. The main parameters of TNT are very similar to those of the Lulin One-meter Telescope
(LOT; Kinoshita et al. \cite{Kino05}), except for differences in the aperture of the main mirror and
the cooling-down mode of the CCD detector.

The CCD detector mounted on TNT is  Princeton Instruments VersArray:1300B
\footnote{$http://www.princetoninstruments.com$}. This is a high-performance, full-frame digital
camera system that utilizes a back-illuminated, scientific-grade CCD. With a 1340$\times$1300 imaging array
(20$\times$20$\mu$m /pixel), this system provides a field-of-view (FOV) of $11.5^{\prime} \times 11.2^{\prime}$
with a spatial resolution of $\sim0.52^{\prime\prime}$ pixel$^{-1}$. The main parameters of
the VersArray:1300B CCD are listed in Table~\ref{Tab:CCD}. It has two readout modes, with the readout time
being about 18 seconds in the slow mode (100 KHz) and about 2 seconds in the fast mode (1 MHz).
There are nominally three 1300B CCDs that have been used on TNT since the start of observation in 2004.
The 1300B-1 CCD had been used before the year 2006, and later replaced by the 1300B-2 CCD during the period
from 5 Jan. to 14 Jun. in year 2006 for maintenance. After that, the 1300B-3 CCD had been used until Sept. 2010
when it was broken, and the 1300B-1 was installed again on TNT as a replacement.

\begin{table}[!htp]
  \caption[]{Parameters of the VersArray:1300B.}
  \label{Tab:CCD}
  \centering
  \begin{tabular}{ll}
                  \hline
                  Features & Specifications \\ \hline
                  Pixel number      & 1340$\times$1300 \\
                  Pixel Size        & 20$\mu$m$\times$20$\mu$m \\
                  Imaging area      & 26.8mm$\times$26mm \\
                  Fill factor       & 100\%  \\
                  AD conversion     & 16 bits \\
                  Scan rates        & 100kHz , 1MHz  \\
                  Full frame readout time & 18s@100kHz, 1.8s@1MHz  \\
                  Read noise        & 2.8$e^{-}$@100kHz, 8$e^{-}$@1MHz  \\
                  Software-selectable gains & 1/2$\times$, 1$\times$, 2$\times$  \\
                  Dark current      & 0.5-1$e^{-}pix^{-1}hr^{-1}$  \\
                  Nonlinearity      & $\leq$ 2\%  \\
                  Cooling medium    & Liquid nitrogen \\
                  Operating temperature & -110$^{\circ}C$ \\
                  Thermostating precision & $\pm$0.05$^{\circ}C$ \\
                  \hline
  \end{tabular}
\end{table}

The filters used on TNT are manufactured by the Custom Scientific, Inc. (USA)
\footnote{$http://www.customscientific.com/astroresearch.html$}, which are the standard
Johnson $UBV$ and Cousin $RI$ system (Bessell \cite{Bessell90}). This has been indicated by a small
color-term correction needed to transform the photometric results from the instrumental system of
TNT to the standard $UBVRI$ system (e.g., Wang et al. \cite{Wang08} \& \cite{Wang09}).

\section{CCD tests}
\label{sect:ccdtest}

\subsection{Bias Level}
A bias level is present in every CCD image, arising from an electronic offset which is added
to the signal of the CCD before being converted to the digital values. Its stability has a
non-negligible effect on the high precision photometry. The bias level has been measured
for all the three CCDs used on TNT in both the slow and fast readout modes. We performed
a continuous 30-hour test of the bias for the 1300B-1 CCD, an 8-hour test for the 1300B-2, and a
7-hour test for the 1300B-3. These results are reported in Table 2.

\begin{figure}[!htp]
   \centering
   \includegraphics[width=90mm]{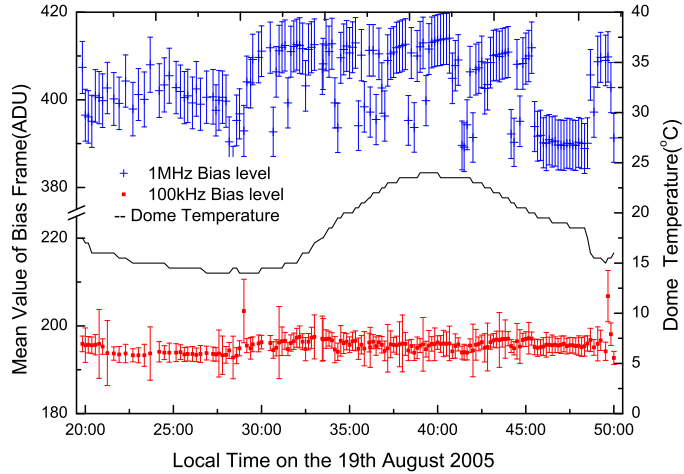}
   \caption{The mean bias level of VersArray: 1300B-1 CCD derived in the slow and fast readout modes
   during a continuous 30 hour period. The solid curve represents the dome temperature during the
   measurements of the bias.}
   \label{Fig:Bias}
\end{figure}

One can see from Fig.~\ref{Fig:Bias} that the mean bias level is related to the readout mode of the CCD.
For the 1300B-1 CCD, this value is about 195 Analog-to-Digital Unit (ADU) for the slow mode and about
400 ADU for the fast mode. We note that the bias measured in the fast readout mode shows some fluctuations,
which might be affected by the ambient environments such as the temperature. Further studies
are needed to clarify this phenomenon. Owing to an instability of the bias level seen in the fast readout
mode, the observers are suggested to take frequent bias frames during observations to achieve
high precision photometry when this mode is used.

\subsection{Gain and Readout Noise}
The gain (G) of a CCD is the ratio between the number of electrons recorded by the CCD chip and
the counts of ADU contained in the CCD image. It is useful to know this value in order to evaluate
the performance of the CCD camera. Knowledge of the gain allows the calculation of the readout noise
(R) and other quantities of the CCD. One can measure the gain of a CCD by comparing the signal
level to the amount of variation in the signal, e.g., in the flatfield images. This works because
the relationship between counts and electrons is different for the signal and the variance.
The gain of a CCD can be determined from the following equation (Howell \cite{Howell00}):
\begin{equation}
G = \frac{({\overline{F1} + \overline{F2}}) - (\overline{B1} +
\overline{B2})}{\sigma^{2}_{F1-F2} - \sigma^{2}_{B1-B2}},
\label{eq:31}
\end{equation}
Where $\overline{F1}$,$\overline{F2}$ are the mean values of different flatfield images, and $\overline{B1}$
,$\overline{B2}$ represent those of the bias images. $\sigma^{2}_{F1-F2}$ and $\sigma^{2}_{B1-B2}$ are
the standard deviation of the difference between two flatfield images and two bias images, respectively.
Subtracting two flatfield images increases the noise by a factor of $\sqrt{2}$. Therefore, the
correlation between the signal $S$ and the noise $N$ can be expressed as
\begin{equation}
N = \sqrt{\frac{S}{G} + (\frac{R}{G})^{2}},
\label{eq:32}
\end{equation}
Here, $R$ is the readout noise.

We took twilight flatfield images in the $B$ band, with the exposure time varying
from 0.1~s to 400~s. At each time of exposure, we took four flatfield images in the slow
readout mode and another four images in the fast readout mode. We then chose 2 better ones and
performed the subtraction to determine the standard deviation $\sigma_{F}$. The noise level
can be obtained by dividing the standard deviation by $\sqrt{2}$. To obtain a mean signal,
we subtracted the combined bias frame from the flatfield images. The noise N and signal S measured
from the flatfield images are used to determine the gain and readout noise through a best fit to
the relation shown by equation (2). We also calculated these two parameters using the task
FINDGAIN in IRAF \footnote{IRAF, the Image Reduction and Analysis Facility, is distributed
by the National Optical Astronomy Observatory, which is operated by the Association of Universities
for Research in Astronomy, Inc. (AURA) under cooperative agreement with the National Science Foundation(NSF).}.
Table ~\ref{Tab:Bias} also lists the resultant gain and the readout noise derived for 1300B-1,2,3 CCDs used
on  TNT.

\begin{table}[!htp]
  \caption[]{Bias, gain, and readout noise determined for the VersArray: 1300B CCD attached to TNT. Two
  readout rates at 100 KHz and 1 MHz are indicated in the brackets of column (2).}
  \label{Tab:Bias}
  \centering
  \begin{tabular}{ll*{3}{@{ }c}}
  \hline \hline
                        & Readout Mode      & 1300B-1        & 1300B-2       & 1300B-3         \\
                        &                   &(before 2006.1) &(2006.1-6)     &(after 2006.6)   \\
  \hline
  Bias(ADU)             & Slow Mode(100kHz) & 195$\pm$3      & 110$\pm$2     & 185$\pm$2         \\
                        & Fast Mode(1MHz)   & 403$\pm$6      & 182$\pm$4     & 213$\pm$5         \\
  \hline \hline
  Readnoise($e^{-}$)    & Slow Mode(100kHz) & 2.75$\pm$0.03  & 2.90$\pm$0.05 & 2.50$\pm$0.21     \\
  (findgain in IRAF)    & Fast Mode(1MHz)   & 9.31$\pm$0.34  & 5.63$\pm$0.08 & 5.94$\pm$0.33     \\
  \hline
  Readnoise($e^{-}$)    & Slow Mode(100kHz) & 2.72$\pm$0.18  & 2.92$\pm$3.90 &  -                \\
  (fit Signal \& Noise) & Fast Mode(1MHz)   & 9.56$\pm$0.50  & 6.16$\pm$3.79 &  -                \\
  \hline \hline
  Gain($e^{-}$/ ADU)    & Slow Mode(100kHz) & 1.96$\pm$0.02  & 1.90$\pm$0.03 & 1.73$\pm$0.09     \\
  (findgain in IRAF)    & Fast Mode(1MHz)   & 2.22$\pm$0.05  & 1.99$\pm$0.03 & 1.81$\pm$0.07     \\
  \hline
  Gain($e^{-}$/ ADU)    & Slow Mode(100kHz) & 1.99$\pm$0.01  & 1.86$\pm$0.02 &  -                \\
  (fit Signal \& Noise) & Fast Mode(1MHz)   & 2.23$\pm$0.01  & 1.99$\pm$0.01 &  -                \\
  \hline \hline
  \end{tabular}
\end{table}

\subsection{Linearity of the CCD Response}
\begin{figure}[!htp]
   \centering
   \includegraphics[width=0.49\textwidth]{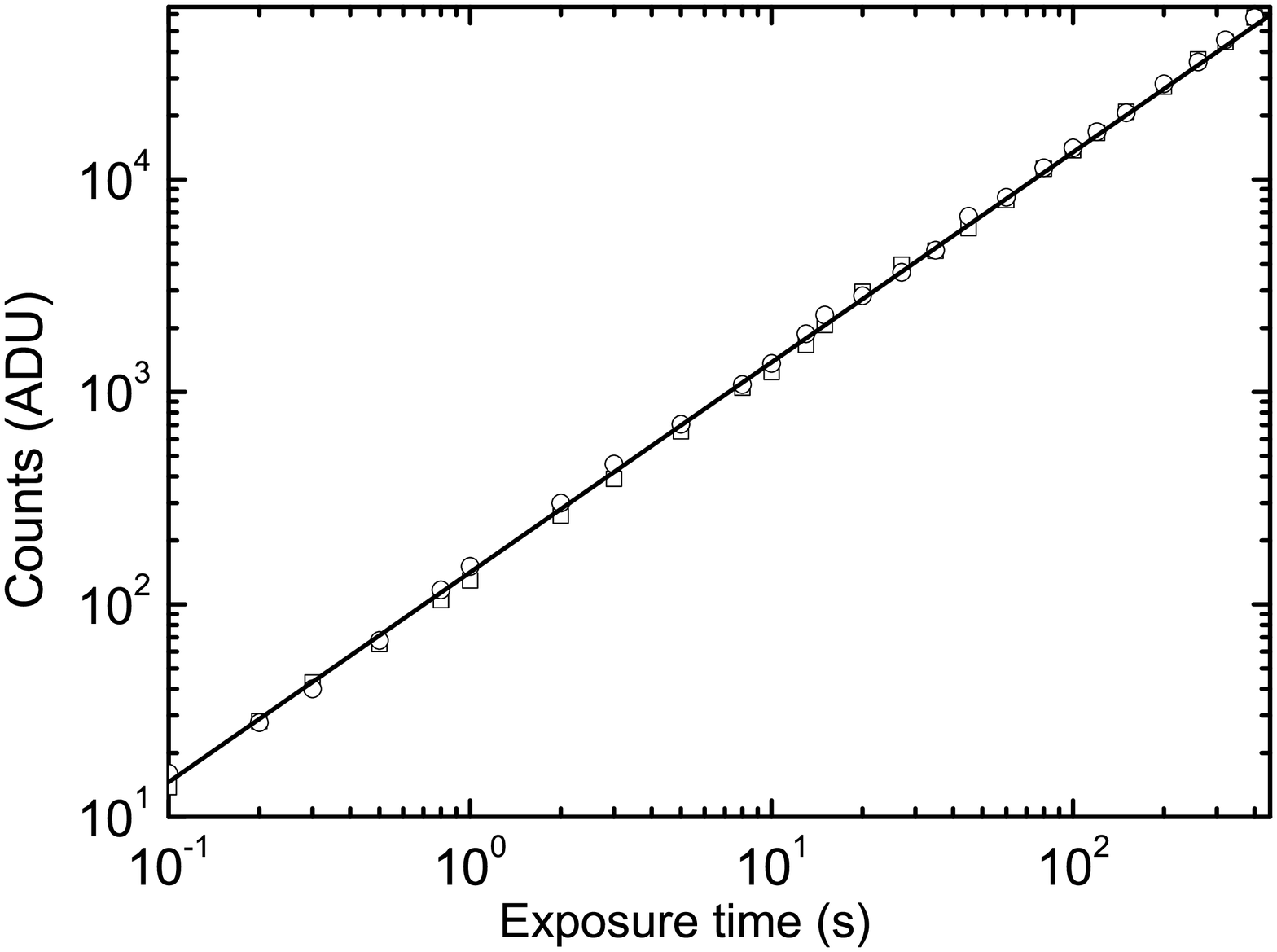}
   \includegraphics[width=0.49\textwidth]{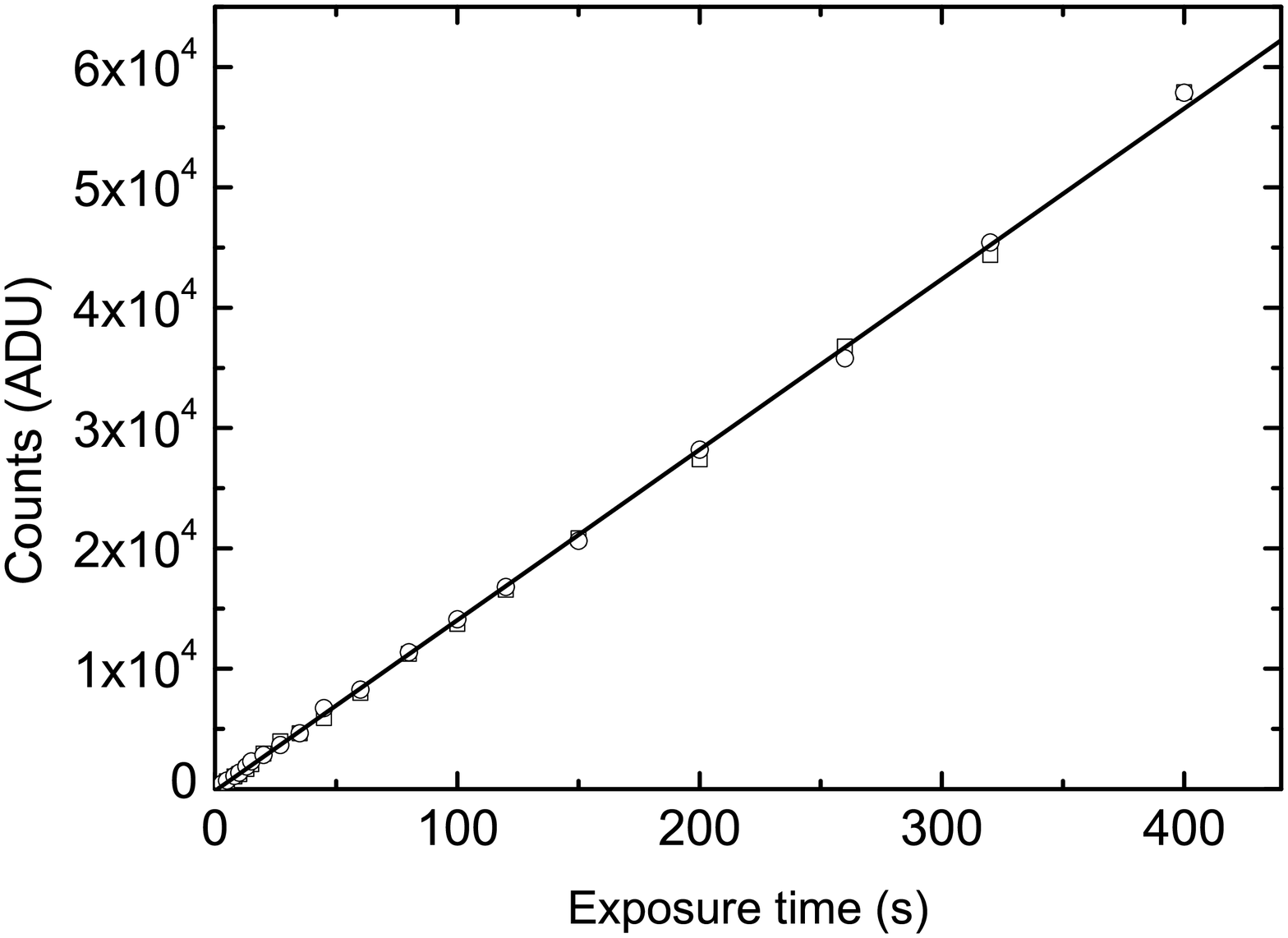}
   \caption{ADU counts of the pixels in the flatfield images as a function of
   the exposure time for the 1300B-1 (left panel) and 1300B-2 (right panel). The counts in the fast mode are
   shown as the squares and those recorded in the slow mode are denoted by the circles. The straight lines show
   linear relationships in both modes.}
   \label{Fig:line}
\end{figure}

One advantage of a modern CCD is its linear response over a large dynamic range. While some pixel values in
the images may be unusable if they are saturated (due to that the charge exceeds the full well capacity) or
are within the nonlinear range. To check the linear response of the VersArray:1300B CCD, we measured the
ADU counts as a function of exposure time using the unfiltered flatfield images. For the 1300B-1 CCD, we use
images taken on August 21, 2005 with the exposure time of 3 $\sim$ 80 s, while 1300B-2 on April 6, 2006 with the
exposure time of 0.1$\sim$400 s. Fig. \ref{Fig:line} shows  a relationship in both the fast and slow modes for the
1300B-1 and the 1300B-2 CCDs respectively. One can see that the linear correlation holds for the pixel
value up to $\sim$50,000 ADU, with the correlation coefficients 0.9998 and 0.9997 respectively.

\subsection{Dark Current}
A routine step of processing the CCD images involves a subtraction of dark current.
Dark current of a CCD usually originates from the collecting of electrons within
the potential well of a pixel in the image, which can become part of the
signal and is indistinguishable from the astronomical photons. It is usually specified
as the number of thermal electrons generated per second per pixel or as the actual
current generated per area of the device. The thermal dark current depends on the CCD
operation temperature (see Figure 3.6 in Howell \cite{Howell00} for a relation between
the dark current and the temperature), which becomes nearly negligible for a
properly cooled CCD.

Dark frames with different integration time (e.g., 600~s and 3600~s) were obtained for
the 1300B-1 CCD to estimate the dark current. We found that the mean dark current generate rate
of the 1300B-1 is about 0.00025$e^{-}s^{-1}pixel^{-1}$ for an operation temperature of
$-$120$^{\circ}C$. It is not surprising that the dark current level of the CCD on  TNT
is much lower than that of the CCD on LOT (i.e., 0.064$e^{-}s^{-1}pixel^{-1}$) since
the working temperature of the latter is much higher with T =$-$50$^{\circ}C$.
With the above dark current generation rate, we estimate that an observation with the
exposure time of 600s will produce a dark current of 0.15$e^{-}pixel^{-1}$. This is far
less than the signal and noise. Hence, we neglect such a minor effect in our image reduction.

\section{PHOTOMETRIC CALIBRATIONS}

\label{sect:phot}

The magnitudes obtained by TNT are the instrumental magnitudes. To compare our photometric
results with those obtained from other instruments, we need to convert our instrumental
magnitudes into the magnitudes defined in the standard $UBVRI$ system. To obtain this
conversion, it is essential to know the transformation equations, which are usually
expressed as:

\begin{equation}
  u =U + Z_U + k'_UX+C_U(U-B),
\label{eq:51}
\end{equation}
\begin{equation}
  b = B + Z_B + k'_BX+C_B(B-V),
\label{eq:52}
\end{equation}
\begin{equation}
  v = V + Z_V + k'_VX+C_V(B-V),
\label{eq:53}
\end{equation}
\begin{equation}
  r = R + Z_R + k'_RX+C_R(V-R),
\label{eq:54}
\end{equation}
\begin{equation}
  i = I + Z_I + k'_IX+C_I(V-I).
\label{eq:55}
\end{equation}
where $ubvri$ are the instrumental magnitudes, $UBVRI$ are the standard magnitudes,
$Z_U, Z_B, Z_V , Z_R, Z_I$ are the zero point magnitudes, $k'_U, k'_B,  k'_V ,  k'_R, k'_I$ are
the first-order extinction coefficients, $C_U, C_B, C_V , C_R, C_I$ are the color terms, and
$X$ is the airmass. The above parameters can be simultaneously determined by observing a series
of Landolt's standard stars covering a certain range of airmass and colors (Landolt \cite{Landolt92}).

Observations of Landolt's standard stars were conducted on twelve photometric nights, spanning the
period from Oct. 2004 to Mar. 2012. Most of these photometric nights are moonless or crescent nights, with
steady and cloudless sky. For a better comparison of these observations obtained at different
time, we divided the photometric nights and the corresponding results into three epochs:
Epoch 1 (2004-2005), Epoch 2 (2006-2007), and Epoch 3 (2011-2012). Table~\ref{Tab:Landolt} lists part
of the Landolt's standard stars that were observed during Epoch 3. The typical exposure time for
these stars is 300s in U, 60s in B, 40s in V, 20s in R, and 20s in I. The photometric data of these
Landolt's standard stars were reduced using the $^{\prime\prime}$apphot$^{\prime\prime}$ package of
IRAF. The deduced coefficients in the transformation equations ((3) - (7)) are shown in Table~\ref{Tab:Tran2}.

\begin{table}[!htp]
   \centering
   \caption[]{The Landolt's standard stars used for the photometric calibration in year 2011 and 2012.}   \label{Tab:Landolt}

   \begin{tabular}{cllcccccc}
   \hline
   Star & $\alpha$(2000) & $\delta$(2000) & V & B-V & U-B & V-R & R-I & V-I          \\
   \hline
     92\_263 & 00:55:40 & +00:36:23 & 11.782 &  1.048 &  0.843 &  0.563 &  0.522 &  1.087 \\
     93\_317 & 01:54:38 & +00:43:11 & 11.546 &  0.488 & -0.055 &  0.293 &  0.298 &  0.592 \\
     94\_251 & 02:57:46 & +00:16:18 & 11.204 &  1.219 &  1.281 &  0.659 &  0.587 &  1.247 \\
     95\_190 & 03:53:13 & +00:16:39 & 12.627 &  0.287 &  0.236 &  0.195 &  0.220 &  0.415 \\
      96\_83 & 04:52:59 & -00:14:22 & 11.719 &  0.179 &  0.202 &  0.093 &  0.097 &  0.190 \\
      97\_75 & 05:57:55 & -00:09:07 & 11.483 &  1.872 &  2.100 &  1.047 &  0.952 &  1.999 \\
     98\_666 & 06:52:10 & -00:23:12 & 12.732 &  0.164 & -0.004 &  0.091 &  0.108 &  0.200 \\
    100\_280 & 08:53:36 & -00:36:24 & 11.799 &  0.494 & -0.002 &  0.295 &  0.291 &  0.588 \\
    101\_413 & 09:56:15 &-00:11:44 & 12.583 &  0.983 &  0.716 &  0.529 &  0.497 &  1.025 \\
    103\_626 & 11:56:47 & -00:21:47 & 11.836 &  0.413 & -0.057 &  0.262 &  0.274 &  0.535 \\
    104\_598 & 12:45:17 & -00:16:41 & 11.479 &  1.106 &  1.050 &  0.670 &  0.546 &  1.215 \\
    105\_815 & 13:40:04 & -00:02:19 & 11.453 &  0.385 & -0.237 &  0.267 &  0.291 &  0.560 \\
   106\_1024 & 14:40:07 & +00:01:31 & 11.599 &  0.332 &  0.085 &  0.196 &  0.195 &  0.390 \\
    107\_484 & 15:40:17 & -00:21:31 & 11.311 &  1.237 &  1.291 &  0.664 &  0.577 &  1.240 \\
    108\_475 & 16:37:00 & -00:35:01 & 11.309 &  1.380 &  1.462 &  0.744 &  0.665 &  1.409 \\
    109\_381 & 17:44:12 & -00:20:55 & 11.730 &  0.704 &  0.225 &  0.428 &  0.435 &  0.861 \\
    110\_280 & 18:43:07 & -00:04:02 & 12.996 &  2.151 &  2.133 &  1.235 &  1.148 &  2.384 \\
   111\_1965 & 19:37:42 & +00:26:30 & 11.419 &  1.710 &  1.865 &  0.951 &  0.877 &  1.830 \\
    112\_250 & 20:42:27 & +00:07:25 & 12.095 &  0.532 & -0.025 &  0.317 &  0.323 &  0.639 \\
    113\_260 & 21:41:49 & +00:23:39 & 12.406 &  0.514 &  0.069 &  0.308 &  0.298 &  0.606 \\
    114\_750 & 22:41:45 & +01:12:30 & 11.916 & -0.041 & -0.354 &  0.027 & -0.015 &  0.011 \\
    RU\_152E & 07:27:25 & -01:58:47 & 12.362 &  0.042 & -0.086 &  0.030 &  0.034 &  0.065 \\
PG1047+003C & 10:50:18 & -00:00:21 & 12.453 &  0.607 & -0.019 &  0.378 &  0.358 &  0.737 \\
 PG2349+002 & 23:51:53 & +00:28:17 & 13.277 & -0.191 & -0.921 & -0.103 & -0.116 & -0.219 \\
    \hline
  \end{tabular}
\end{table}

\begin{table}[!htp]
  \caption[]{Transformation coefficients of zero-point magnitudes, first-order atmospheric extinction
coefficients, and color terms in the $UBVRI$ bands, derived from the calibration data of 12 photometric nights.}
  \label{Tab:Tran2}
  \centering
  \begin{tabular}{*{6}{c}}
  \hline \hline
  Date(ymd) & $Z_U$ & $Z_B$ & $Z_V$ & $Z_R$ & $Z_I$ \\
  \hline
  20041026    &  1.162$\pm$0.145 & -1.317$\pm$0.046 & -1.711$\pm$0.036 & -1.671$\pm$0.039 & -0.872$\pm$0.032 \\
  20041127    &                  & -1.356$\pm$0.028 & -1.687$\pm$0.021 & -1.604$\pm$0.020 & -0.875$\pm$0.017 \\
  20050902    &                  & -1.023$\pm$0.017 & -1.502$\pm$0.011 & -1.552$\pm$0.016 & -0.910$\pm$0.024 \\
  \hline
  Epoch1 mean &  1.162$\pm$0.048 & -1.232$\pm$0.019 & -1.633$\pm$0.014 & -1.609$\pm$0.016 & -0.886$\pm$0.014 \\
  \hline
  20061221    & -0.186$\pm$0.046 & -1.854$\pm$0.024 & -2.040$\pm$0.021 & -2.155$\pm$0.020 & -1.692$\pm$0.029 \\
  20070107    &  0.032$\pm$0.045 & -1.798$\pm$0.034 & -1.978$\pm$0.028 & -2.113$\pm$0.026 & -1.573$\pm$0.021 \\
  20070111    & -0.042$\pm$0.061 & -1.770$\pm$0.025 & -1.935$\pm$0.020 & -2.048$\pm$0.017 & -1.586$\pm$0.024 \\
  20071212    & -0.123$\pm$0.061 & -1.837$\pm$0.032 & -2.025$\pm$0.032 & -2.085$\pm$0.033 & -1.659$\pm$0.028 \\
  \hline
  Epoch2 mean & -0.080$\pm$0.027 & -1.815$\pm$0.014 & -1.995$\pm$0.013 & -2.100$\pm$0.012 & -1.628$\pm$0.012 \\
  \hline
  20111024    &  5.650$\pm$0.369 &  3.535$\pm$0.082 &  3.210$\pm$0.036 &  3.231$\pm$0.019 &  3.691$\pm$0.068 \\
  20111223    &  5.524$\pm$0.105 &  3.620$\pm$0.016 &  3.259$\pm$0.010 &  3.304$\pm$0.008 &  3.729$\pm$0.036 \\
  20111231    &  5.900$\pm$0.066 &  3.646$\pm$0.049 &  3.214$\pm$0.071 &  3.350$\pm$0.040 &  3.756$\pm$0.045 \\
  20120306    &                  &  3.900$\pm$0.126 &  3.650$\pm$0.110 &  3.650$\pm$0.119 &  4.200$\pm$0.117 \\
  20120327    &                  &  3.982$\pm$0.035 &  3.566$\pm$0.027 &  3.595$\pm$0.029 &  4.047$\pm$0.021 \\
  \hline
  Epoch3 mean &  5.691$\pm$0.078 &  3.737$\pm$0.033 &  3.380$\pm$0.028 &  3.426$\pm$0.026 &  3.885$\pm$0.030 \\
  \hline
  \hline
  Date(ymd) & $k'_U$ & $k'_B$ & $k'_V$ & $k'_R$ & $k'_I$ \\
  \hline
  20041026    &  0.699$\pm$0.095 &  0.311$\pm$0.030 &  0.211$\pm$0.023 &  0.153$\pm$0.025 &  0.069$\pm$0.020 \\
  20041127    &                  &  0.306$\pm$0.020 &  0.201$\pm$0.014 &  0.121$\pm$0.015 &  0.089$\pm$0.011 \\
  20050902    &                  &  0.272$\pm$0.011 &  0.184$\pm$0.008 &  0.149$\pm$0.011 &  0.092$\pm$0.016 \\
  \hline
  Epoch1 mean &  0.699$\pm$0.032 &  0.296$\pm$0.012 &  0.199$\pm$0.009 &  0.141$\pm$0.010 &  0.083$\pm$0.009 \\
  \hline
  20061221    &  0.648$\pm$0.027 &  0.295$\pm$0.014 &  0.201$\pm$0.012 &  0.152$\pm$0.011 &  0.099$\pm$0.016 \\
  20070107    &  0.548$\pm$0.028 &  0.295$\pm$0.020 &  0.214$\pm$0.016 &  0.175$\pm$0.015 &  0.074$\pm$0.012 \\
  20070111    &  0.644$\pm$0.040 &  0.332$\pm$0.016 &  0.220$\pm$0.013 &  0.158$\pm$0.011 &  0.093$\pm$0.016 \\
  20071212    &  0.709$\pm$0.038 &  0.307$\pm$0.020 &  0.221$\pm$0.020 &  0.158$\pm$0.021 &  0.097$\pm$0.018 \\
  \hline
  Epoch2 mean &  0.637$\pm$0.017 &  0.307$\pm$0.009 &  0.214$\pm$0.008 &  0.161$\pm$0.008 &  0.091$\pm$0.008 \\
  \hline
  20111024    &  0.548$\pm$0.242 &  0.310$\pm$0.055 &  0.180$\pm$0.021 &  0.130$\pm$0.021 &  0.104$\pm$0.053 \\
  20111223    &  0.602$\pm$0.147 &  0.309$\pm$0.009 &  0.220$\pm$0.006 &  0.161$\pm$0.009 &  0.089$\pm$0.024 \\
  20111231    &  0.510$\pm$0.053 &  0.360$\pm$0.038 &  0.316$\pm$0.026 &  0.215$\pm$0.046 &  0.120$\pm$0.030 \\
  20120306    &                  &  0.447$\pm$0.086 &  0.242$\pm$0.074 &  0.180$\pm$0.078 &  0.031$\pm$0.079 \\
  20120327    &                  &  0.314$\pm$0.025 &  0.222$\pm$0.020 &  0.153$\pm$0.020 &  0.083$\pm$0.016 \\
  \hline
  Epoch3 mean &  0.553$\pm$0.058 &  0.348$\pm$0.022 &  0.236$\pm$0.017 &  0.168$\pm$0.019 &  0.085$\pm$0.021 \\
  \hline
  \hline

  Date(ymd) & $C_U$ & $C_B$ & $C_V$ & $C_R$ & $C_I$ \\
  \hline
  20041026    & -0.301$\pm$0.023 & -0.190$\pm$0.011 &  0.077$\pm$0.009 &  0.135$\pm$0.017 & -0.043$\pm$0.007 \\
  20041127    &                  & -0.165$\pm$0.005 &  0.083$\pm$0.004 &  0.146$\pm$0.006 & -0.036$\pm$0.003 \\
  20050902    &                  & -0.229$\pm$0.005 &  0.067$\pm$0.004 &  0.085$\pm$0.009 & -0.024$\pm$0.007 \\
  \hline
  Epoch1 mean & -0.301$\pm$0.008 & -0.195$\pm$0.004 &  0.076$\pm$0.003 &  0.122$\pm$0.007 & -0.034$\pm$0.003 \\
  \hline
  20061221    & -0.132$\pm$0.016 & -0.132$\pm$0.008 &  0.086$\pm$0.007 &  0.110$\pm$0.011 & -0.037$\pm$0.009 \\
  20070107    & -0.107$\pm$0.013 & -0.128$\pm$0.011 &  0.076$\pm$0.009 &  0.106$\pm$0.016 & -0.035$\pm$0.007 \\
  20070111    & -0.136$\pm$0.011 & -0.134$\pm$0.006 &  0.080$\pm$0.005 &  0.105$\pm$0.007 & -0.040$\pm$0.005 \\
  20071212    & -0.124$\pm$0.016 & -0.133$\pm$0.007 &  0.079$\pm$0.007 &  0.101$\pm$0.013 & -0.038$\pm$0.006 \\
  \hline
  Epoch2 mean & -0.125$\pm$0.007 & -0.132$\pm$0.004 &  0.080$\pm$0.004 &  0.106$\pm$0.006 & -0.038$\pm$0.004 \\
  \hline
  20111024    & -0.316$\pm$0.063 & -0.144$\pm$0.011 &  0.068$\pm$0.007 &  0.108$\pm$0.007 & -0.026$\pm$0.011 \\
  20111223    & -0.218$\pm$0.027 & -0.149$\pm$0.002 &  0.064$\pm$0.004 &  0.095$\pm$0.004 & -0.023$\pm$0.006 \\
  20111231    & -0.367$\pm$0.034 & -0.146$\pm$0.010 &  0.064$\pm$0.009 &  0.076$\pm$0.015 & -0.025$\pm$0.007 \\
  20120306    &                  & -0.164$\pm$0.035 &  0.071$\pm$0.029 &  0.088$\pm$0.055 & -0.015$\pm$0.031 \\
  20120327    &                  & -0.155$\pm$0.008 &  0.062$\pm$0.007 &  0.083$\pm$0.014 & -0.033$\pm$0.005 \\
  \hline
  Epoch3 mean & -0.300$\pm$0.015 & -0.152$\pm$0.008 &  0.066$\pm$0.006 &  0.090$\pm$0.012 & -0.024$\pm$0.007 \\
  \hline
  \hline
  \end{tabular}
\end{table}

The photometric data obtained at different epochs seem to give a similar mean value of the relevant
coefficients except for the magnitude zeropoints. The large difference in the magnitude zeropoint
between Epoch 3 and the other two epochs are primarily related to the specific definition of the
magnitude zeropoint, e.g., with an offset of 5.0 mag in all of the $UBVRI$ bands.
The mean atmospheric extinction coefficients\footnote{The extinction coefficients are in unit of magnitude per airmass.}
at Xinglong Observatory, obtained with the most recent data (e.g., Epoch 3), are 0.55$\pm$0.06 in $U$, 0.35$\pm$0.02 in B,
0.24$\pm$0.02 in V, 0.17$\pm$0.02 in R, and 0.09$\pm$0.02 in I, respectively. Recently, Zhou et al. (\cite{Zhou09}) also
examined the atmospheric extinction at Xinglong based on the observations with the 85-cm telescope. Their studies show
that the first-order atmospheric extinction coefficient in the BVRI bands are 0.33$\pm$0.01, 0.24$\pm$0.01, 0.20$\pm$0.01,
and 0.07$\pm$0.01, respectively, which are consistent with ours within the quoted errors. In Table~\ref{Tab:extin}, we also
compared our results with two earlier estimates for the site given by Shi et al. (\cite{Shi98}).

\begin{table}[!htp]
  \caption[]{Atmospheric extinction coefficients at Xinglong Observatory.}
  \label{Tab:extin}
  \begin{center}
  \begin{tabular}{*{6}{c}}
  \hline
   Year      & $k'_B$ & $k'_V$ & $k'_R$ & $k'_I$ & References\\
   \hline
   2011-2012 & 0.348$\pm$0.022 &  0.236$\pm$0.017 &  0.168$\pm$0.019 &  0.085$\pm$0.021 & 1\\
   2008      & 0.330$\pm$0.007 &  0.242$\pm$0.005 &  0.195$\pm$0.004 &  0.066$\pm$0.003 & 2\\
   2006-2007 & 0.307$\pm$0.009 &  0.214$\pm$0.008 &  0.161$\pm$0.008 &  0.091$\pm$0.008 & 1\\
   2004-2005 & 0.296$\pm$0.012 &  0.199$\pm$0.009 &  0.141$\pm$0.010 &  0.083$\pm$0.009 & 1\\
   1995      & 0.35            &  0.20            &  0.18            &  0.16            & 3\\
   1989      & 0.31            &  0.22            &  0.14            &  0.10            & 3\\
  \hline
   \end{tabular}
\end{center}
REFERENCES:~~1. this paper  ~~2. Zhou et al. (\cite{Zhou09})  ~~3. Shi et al. (\cite{Shi98})
  \end{table}

\begin{figure}[!htp]
   \centering
   \includegraphics[width=0.49\textwidth]{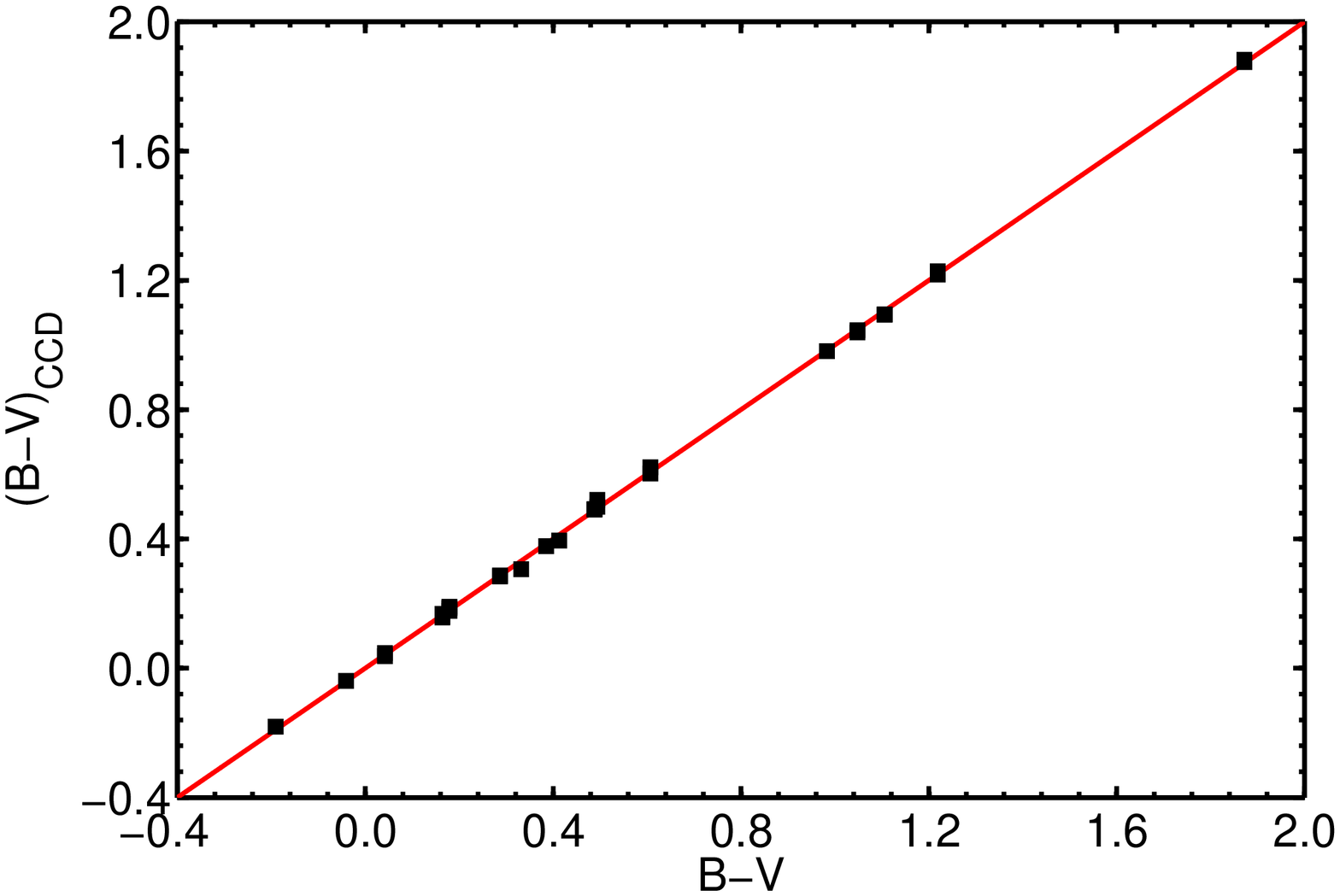}
   \includegraphics[width=0.49\textwidth]{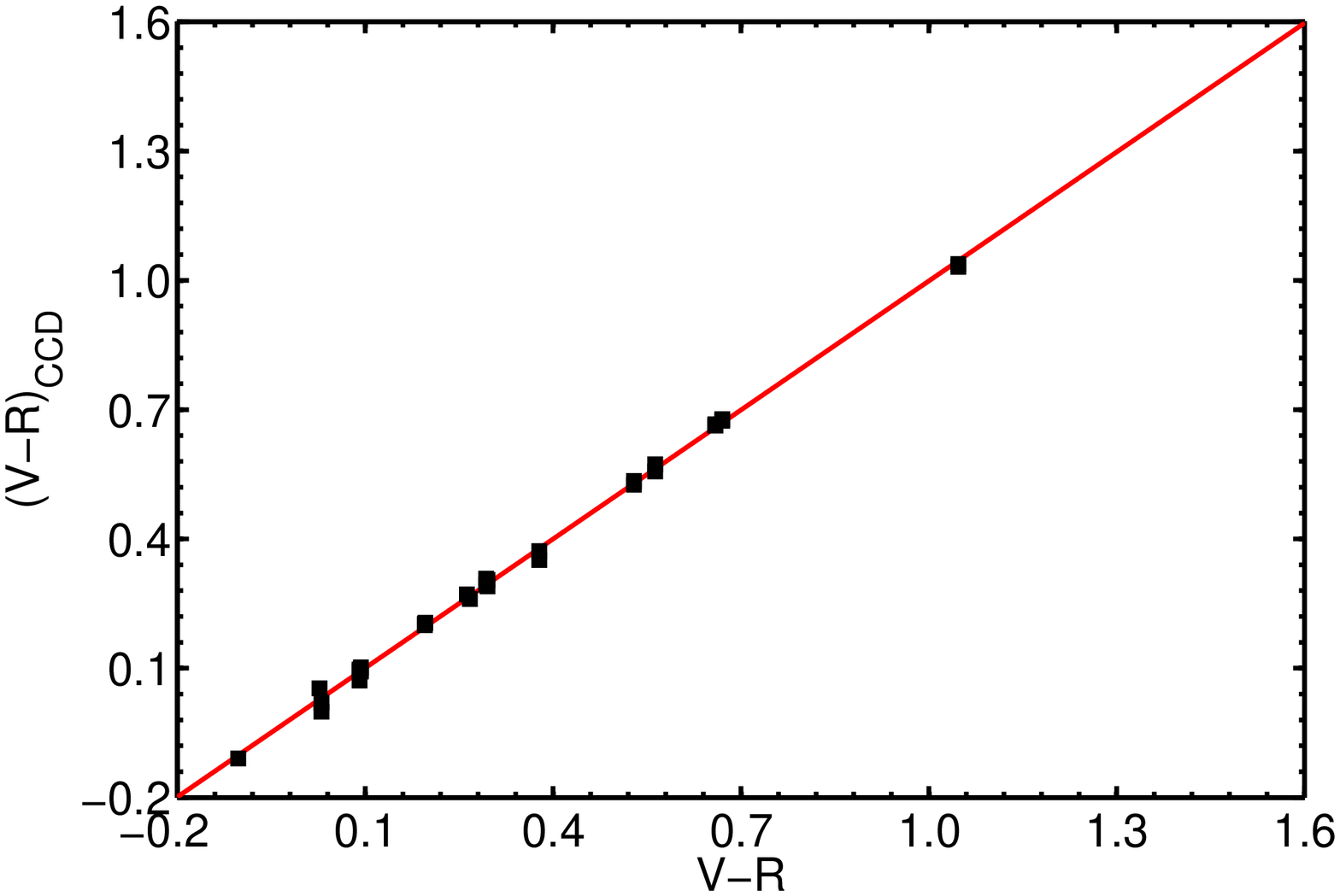}
   \includegraphics[width=0.49\textwidth]{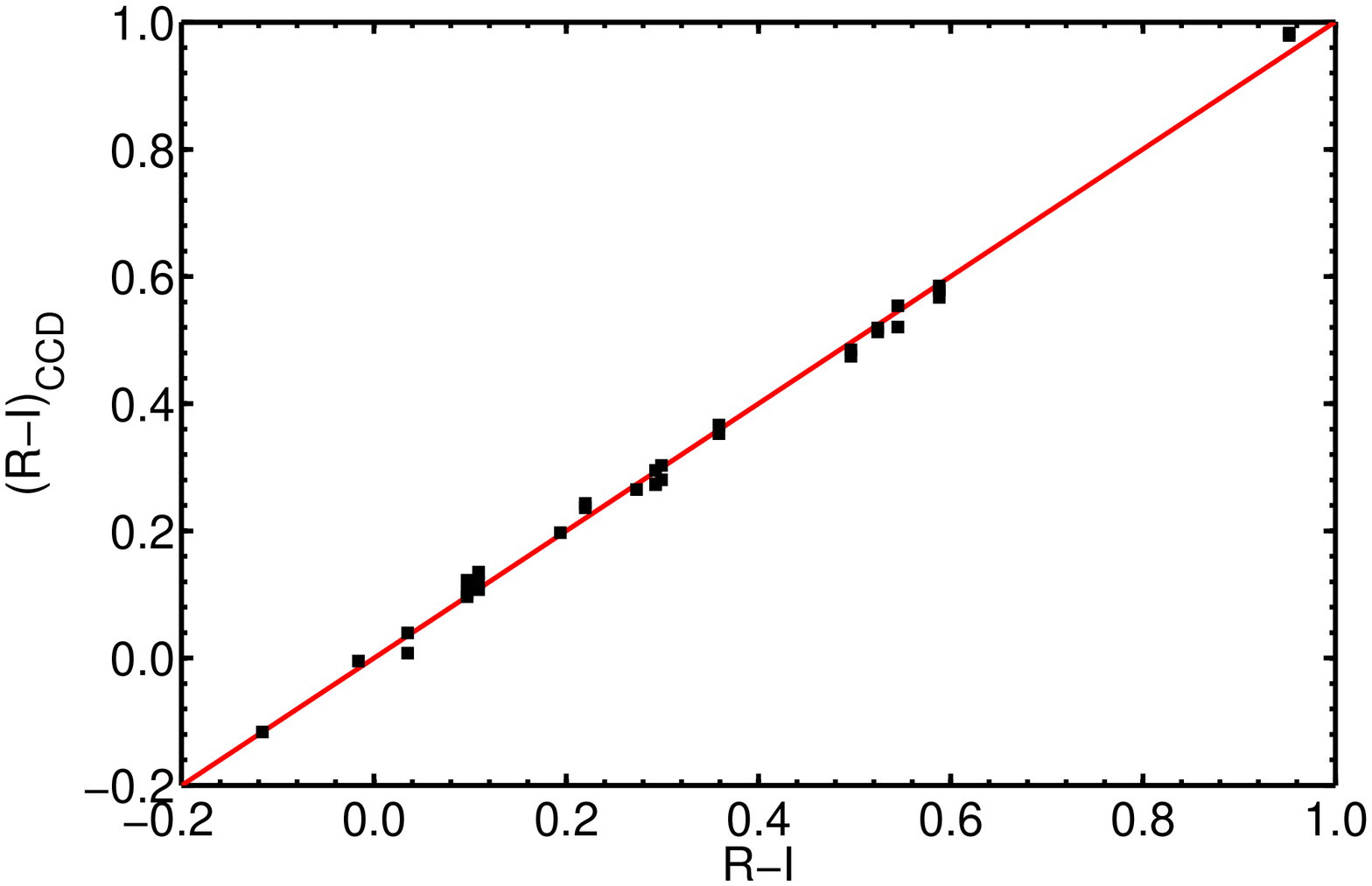}
   \includegraphics[width=0.49\textwidth]{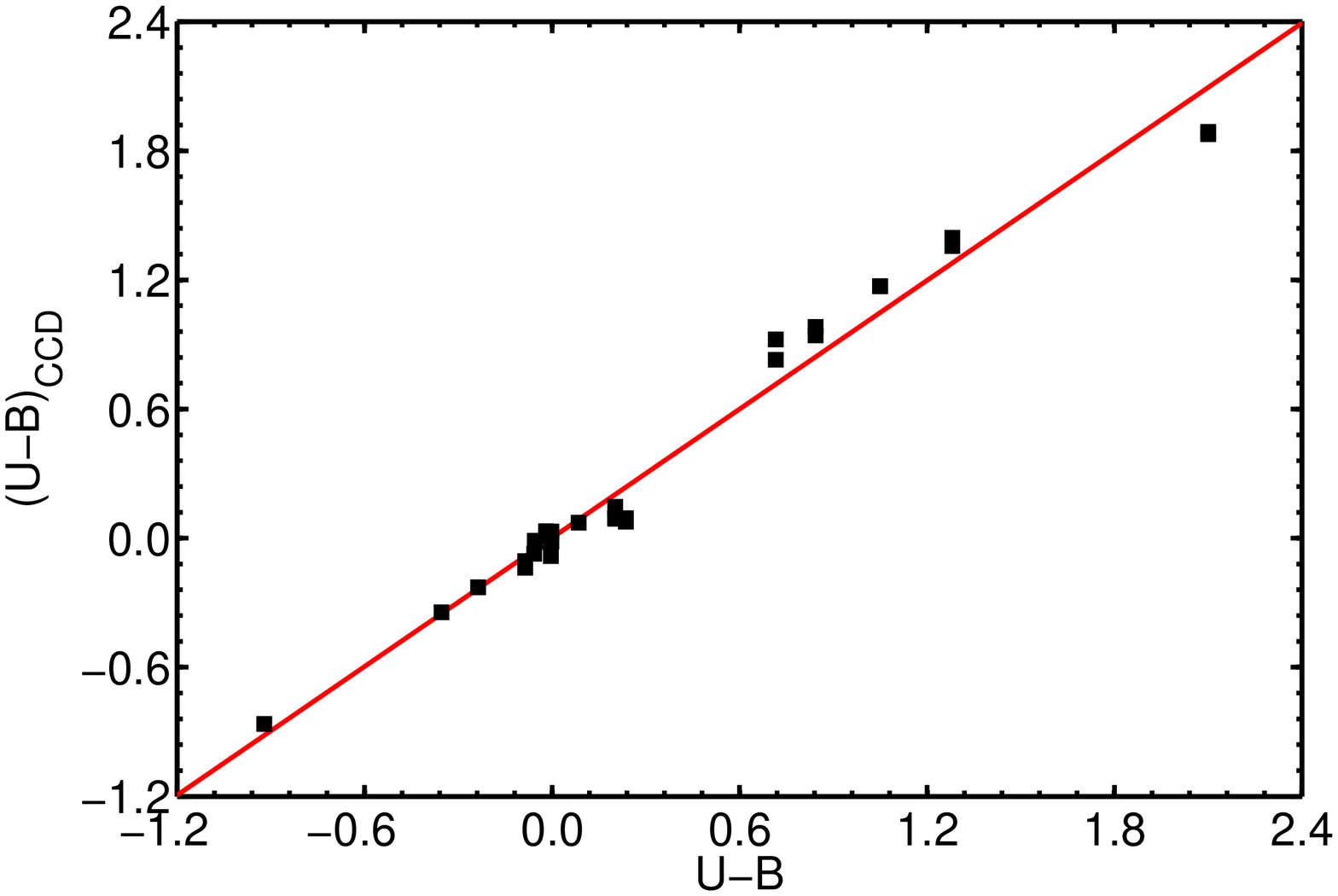}
   \caption{The relationship between the Landolt (1992) color indices and the colors deduced from
   the transformation equations ((3)-(7)). The data taken on 31 Dec., 2011 are used for the plot.}
\label{Fig:magcp}
\end{figure}

The color terms determined at the above three epochs are generally in accordance with each other,
except in the $U$ band where the variation is likely related to the change of the CCD that directly determines
the quantum efficiency and hence the profile of the instrumental response curve.
The 1300B-1 CCD was used during the periods over Epoch 1 and Epoch 3, and the corresponding photometric system has a
larger U-band color term; while the 1300B-2 CCD photometric system has a smaller value. Figure~\ref{Fig:magcp} shows
the correlations between the Landolt colors (Landolt 1992) and the instrument colors of TNT transformed by
equations ((3)-(7)). The Landolt standard stars observed on 31 Dec, 2011 are used for the plot.
Fitting those data points in a linear fashion yields a slope that is very close to 1.0, with an rms $<$ 0.1
mag in different filters\footnote{The scatter in U is slightly larger ($\sim$0.58 mag) due to relatively lower quality data.}.
This means that transformation from the photometric system of  TNT to the Johnson-Cousion standard photometric system
can be well established.

Note that the above color coefficients are obtained with normal stars with the $B - V$ color
ranging from $-$0.3 mag to +2.2 mag, and may not account for the whole photometric differences between
the instrumental magnitudes and the standard Johnson-Cousion magnitudes for some variable sources
such as SNe and GRBs because of their peculiar spectral shapes and features. Besides the color
term correction, additional corrections such as the S-corrections (Stritzinger et al. 2002) are
usually required for precise photometry of these objects.

\section{SYSTEM PERFORMANCE} \label{sect:syst}

\subsection{System Efficiency}

Using the photometric observations of the Landolt's standard stars,
we could also estimate the total throughput of the overall observation
system. This involves the filter response, the atmospheric transmission,
the telescope optics and the detector quantum efficiency. Following the
descriptions by Kinoshita et al. (\cite{Kino05}) (see their Equations (15)-(18)),
we computed the throughput efficiency of 1300B-1 CCD for the TNT observations. The
results in different bands are summarized in Table~\ref{Tab:Eff}, with higher
throughput efficiency in V and R bands. This is in accordance with the CCD
quantum efficiency light curves provided by the manufacturers.

\begin{table}[!htp]
  \caption[]{The total throughout of the TNT photometric system for the $UBVRI$ bands,
  including telescope optics, filter transmittance, and detector quantum efficiency. }
  \label{Tab:Eff}
  \centering
  \begin{tabular}{cccccc}
  \hline
  Band & U & B & V & R & I \\

  Throughput & 9.5\%& 12.1\%& 24.7\%& 36.4\%& 13.6\%  \\
  \hline
  \end{tabular}
\end{table}

\subsection{Sky Background Brightness}
As a byproduct of our photometric calibrations, we could also estimate the brightness of the night
sky based on the flux of the sky background. The instrumental magnitudes were converted into the
standard system with the transformation equations (\ref{eq:51}) - (\ref{eq:55}) and the coefficients
shown in Table 3. We did not consider the effects caused by the difference in the airmass and the
direction of the sky area\footnote{Xinglong Observatory is located on the northeast 115~km of Beijing,
northeast 200~km of Tianjin, and east 7~km of Xinglong county. The city light from Beijing, Tianjin, 
and Xinglong can contribute significantly to the western sky background around Xinglong Observatory.}. 
As the sky background emission is affected significantly by the moon phase, we divided the twelve-night 
data into two groups: moonlit and moonless nights. During the moonless nights (2004$\sim$2007), the sky brightness
was estimated to $\sim$21.8 mag in $U$, $\sim$21.7 mag in $B$, $\sim$21.2 mag in $V$, $\sim$20.5 mag
in $R$, and $\sim$19.1 mag in $I$, respectively.These values are generally consistent with the estimates
obtained in year 1989 and year 1995 (Shi et al. \cite{Shi98})(see also Table~\ref{Tab:sky}). The value
in $V$ is also consistent with the mean value of the moonlight-corrected sky brightness derived from
the BATC data(Liu et al. \cite{Liu03}). In year 2011, however, the sky brightness is found to be
$\sim$21.3 mag in $U$,$\sim$20.9 mag in $B$, $\sim$20.0 mag in $V$, $\sim$19.3 mag in $R$, and
$\sim$18.1 mag in $I$, respectively. These values are apparently brighter than those obtained a couple
of years ago, indicating that the sky background at Xinglong Observatory is becoming worse in recent
years. This is perhaps related to the contamination of the city light of Beijing, Tianjin and Xinglong.

\begin{table}[!htp]
  \caption[]{The night sky background brightness at Xinglong Observatory.
  The brightness is expressed in the unit of mag arcsec$^{-2}$.}
  \label{Tab:sky}
  \centering
  \begin{tabular}{*{7}{@{ }c}}
  \hline
   Date(ymd) & U & B & V & R & I \\
   \hline
   20041026      & 19.831$\pm$0.261 &    19.434$\pm$0.079 &  19.211$\pm$0.060 &  18.961$\pm$0.441 &  18.321$\pm$0.180 \\
   20041127      &                  &    19.377$\pm$0.044 &  19.467$\pm$0.032 &  19.226$\pm$0.186 &  18.672$\pm$0.081 \\
   20070107      & 19.388$\pm$0.186 &    19.725$\pm$0.050 &  19.678$\pm$0.042 &  19.471$\pm$0.393 &  18.650$\pm$0.155 \\
   20120306      &                  &    17.025$\pm$0.015 &  16.857$\pm$0.015 &  16.525$\pm$0.019 &  16.212$\pm$0.018 \\
 \hline
   moonlit  above  \\
  \hline
  20050902       &                  &    21.724$\pm$0.024 &  21.451$\pm$0.016 &  20.914$\pm$0.243 &  19.411$\pm$0.180 \\
  20061221       & 22.312$\pm$0.287 &    22.195$\pm$0.035 &  21.416$\pm$0.031 &  20.474$\pm$0.309 &  18.908$\pm$0.225 \\
  20070111       & 21.709$\pm$0.148 &    21.606$\pm$0.039 &  21.083$\pm$0.031 &  20.358$\pm$0.197 &  19.079$\pm$0.143 \\
  20071212       & 21.405$\pm$0.269 &    21.279$\pm$0.049 &  20.764$\pm$0.049 &  20.142$\pm$0.363 &  18.896$\pm$0.165 \\
  20111024       & 22.352$\pm$0.465 &    21.129$\pm$0.147 &  19.847$\pm$0.086 &  19.279$\pm$0.101 &  18.117$\pm$0.062 \\
  20111223       & 21.571$\pm$0.147 &    21.270$\pm$0.091 &  20.286$\pm$0.063 &  19.486$\pm$0.061 &  17.970$\pm$0.035 \\
  20111231       & 20.655$\pm$0.077 &    20.490$\pm$0.063 &  20.110$\pm$0.042 &  19.382$\pm$0.045 &  18.100$\pm$0.026 \\
  20120327       &                  &    20.995$\pm$0.061 &  20.106$\pm$0.043 &  19.371$\pm$0.044 &  18.109$\pm$0.028 \\
  \hline
  moonless  above \\
  \hline
   1989 10      &                  & 22.15 & 21.04 & 20.25 & 18.77 \\
   1995 10      &                  & 21.81 & 20.60 & 20.22 & 18.46 \\
  \hline
  \end{tabular}
\end{table}

\subsection{Limiting Magnitude and Photometric Precision}

We estimated the limiting magnitudes of the TNT photometric system as well.
We used the equation below (Howell \cite{Howell00}) to perform our calculation:
\begin{equation}
  \frac{S}{N}=\frac{N_{star}}{\sqrt{N_{star}+n_{pix}(N_{sky}+N_{dark}+N^{2}_{readout})}}.
\label{eq:66}
\end{equation}
$N_{star}$ is the total number of photons collected from the targets. $N_{sky}$ is the total number
of photons per pixel from the sky background. $N_{dark}$ is the dark current per pixel from thermal electrons.
$N_{readout}$ is the readout noise estimated in Section 2.2. $n_{pix}$ is the number of pixels under considerations
for the calculation. The dark current electrons are neglected here.

The limiting magnitudes derived on the moonless nights with an aperture size of 6 arcsec in a slow readout mode
are listed in Table~\ref{Tab:limit}, with a mean value of U$\sim$19.2 mag, B$\sim$19.0 mag, V$\sim$18.8 mag,
R$\sim$18.7 mag, and I$\sim$18.2 mag, respectively, for a signal-to-noise ratio (SNR) of 100 and an integration
time of 300~s. Compared with detection limit for the 85-cm telescope obtained by Zhou et al. (\cite{Zhou09})
(see their Table D.2), TNT seems to go slightly deeper, e.g. 18.8 mag vs. 18.2 mag in V for a 300-s exposure, given a similar exposure time and SNR.

\begin{table}[!htp]
  \caption[]{Limiting magnitudes (for a signal-to-noise ratio of 100) derived in the slow
  mode using the data taken on the moonlit and moonless nights with an exposure time of 300~s and
  a photometric aperture of 6 arcsec.}
  \label{Tab:limit}
  \centering
  \begin{tabular}{*{7}{@{ }c}}
  \hline
  Date(ymd) & U & B & V & R & I \\
  \hline
  20041026      & 18.280$\pm$0.122 &    18.091$\pm$0.038 &  17.984$\pm$0.029 &  17.863$\pm$0.215 &  17.551$\pm$0.089 \\
  20041127      &                  &    18.064$\pm$0.021 &  18.107$\pm$0.015 &  17.992$\pm$0.090 &  17.723$\pm$0.039 \\
  20070107      & 18.069$\pm$0.089 &    18.230$\pm$0.024 &  18.207$\pm$0.020 &  18.109$\pm$0.189 &  17.712$\pm$0.076 \\
  20111006      &                  &    17.795$\pm$0.021 &  17.666$\pm$0.021 &  17.889$\pm$0.020 &  17.936$\pm$0.021 \\
  20120306      &                  &    17.577$\pm$0.007 &  17.527$\pm$0.007 &  17.744$\pm$0.009 &  17.785$\pm$0.009 \\
 \hline
  moonlit mean  & 18.175$\pm$0.030 &    17.951$\pm$0.011 &  17.898$\pm$0.009 &  17.919$\pm$0.060 &  17.741$\pm$0.025 \\
  \hline
  20050902      &                  &    19.108$\pm$0.009 &  19.000$\pm$0.007 &  18.773$\pm$0.107 &  18.080$\pm$0.087 \\
  20061221      & 19.321$\pm$0.092 &    19.280$\pm$0.012 &  18.985$\pm$0.013 &  18.577$\pm$0.142 &  17.838$\pm$0.109 \\
  20070111      & 19.102$\pm$0.056 &    19.061$\pm$0.015 &  18.846$\pm$0.013 &  18.525$\pm$0.091 &  17.920$\pm$0.069 \\
  20071212      & 18.981$\pm$0.107 &    18.929$\pm$0.021 &  18.707$\pm$0.022 &  18.425$\pm$0.170 &  17.832$\pm$0.080 \\
  20111024      & 19.353$\pm$0.017 &    19.001$\pm$0.015 &  18.697$\pm$0.016 &  18.761$\pm$0.016 &  18.452$\pm$0.016 \\
  20111223      & 19.257$\pm$0.008 &    19.078$\pm$0.008 &  18.865$\pm$0.009 &  18.855$\pm$0.009 &  18.410$\pm$0.010 \\
  20111231      & 19.116$\pm$0.006 &    18.902$\pm$0.008 &  18.772$\pm$0.007 &  18.815$\pm$0.007 &  18.474$\pm$0.007 \\
  20120327      &                  &    19.104$\pm$0.029 &  18.902$\pm$0.021 &  18.919$\pm$0.021 &  18.582$\pm$0.013 \\
  \hline
  moonless mean & 19.215$\pm$0.019 &    19.058$\pm$0.006 &  18.847$\pm$0.005 &  18.706$\pm$0.033 &  18.199$\pm$0.022 \\
  \hline
  \end{tabular}
\end{table}

\begin{figure}[]
   \centering
   \includegraphics[width=0.49\textwidth]{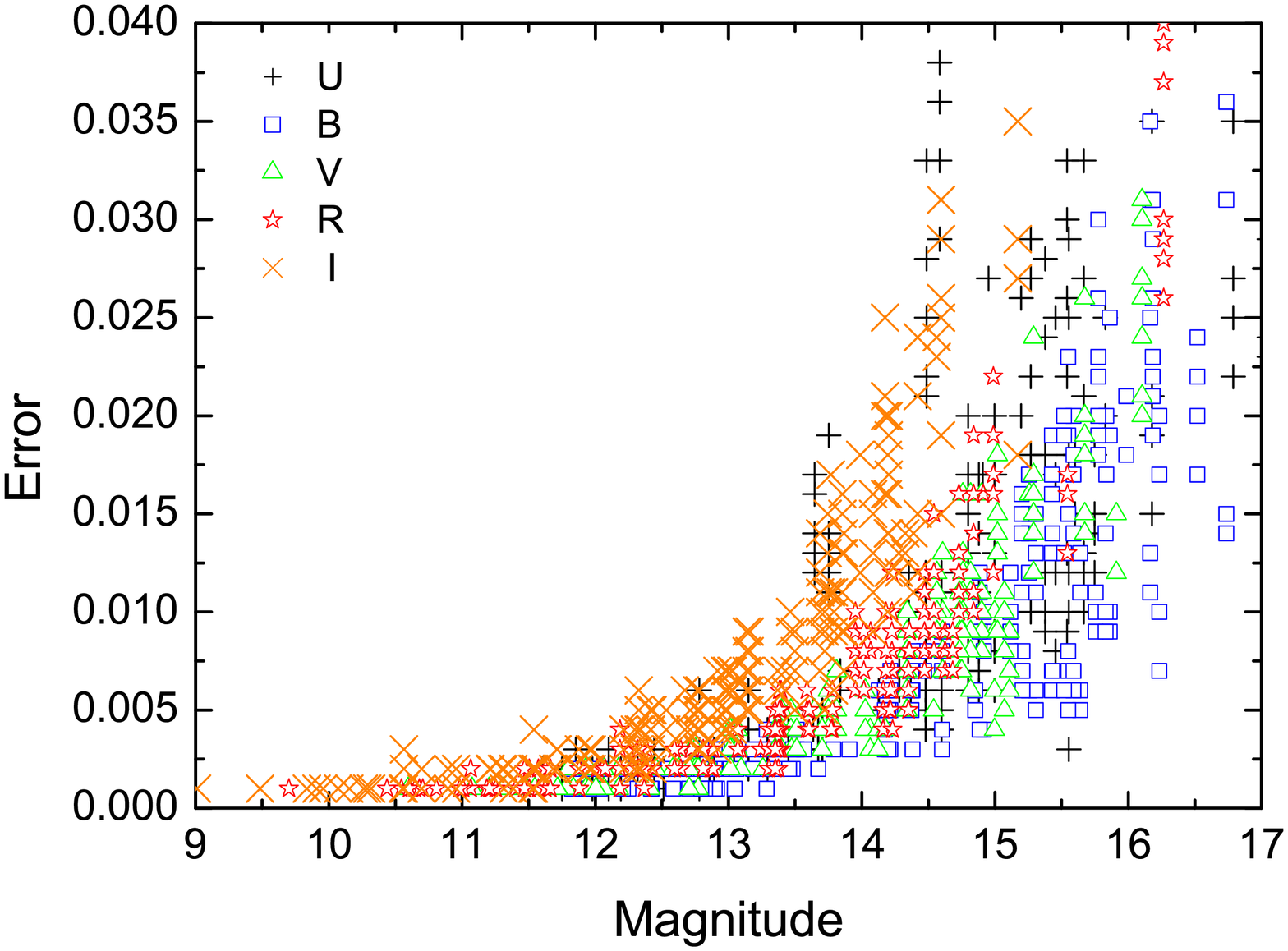}
   \includegraphics[width=0.49\textwidth]{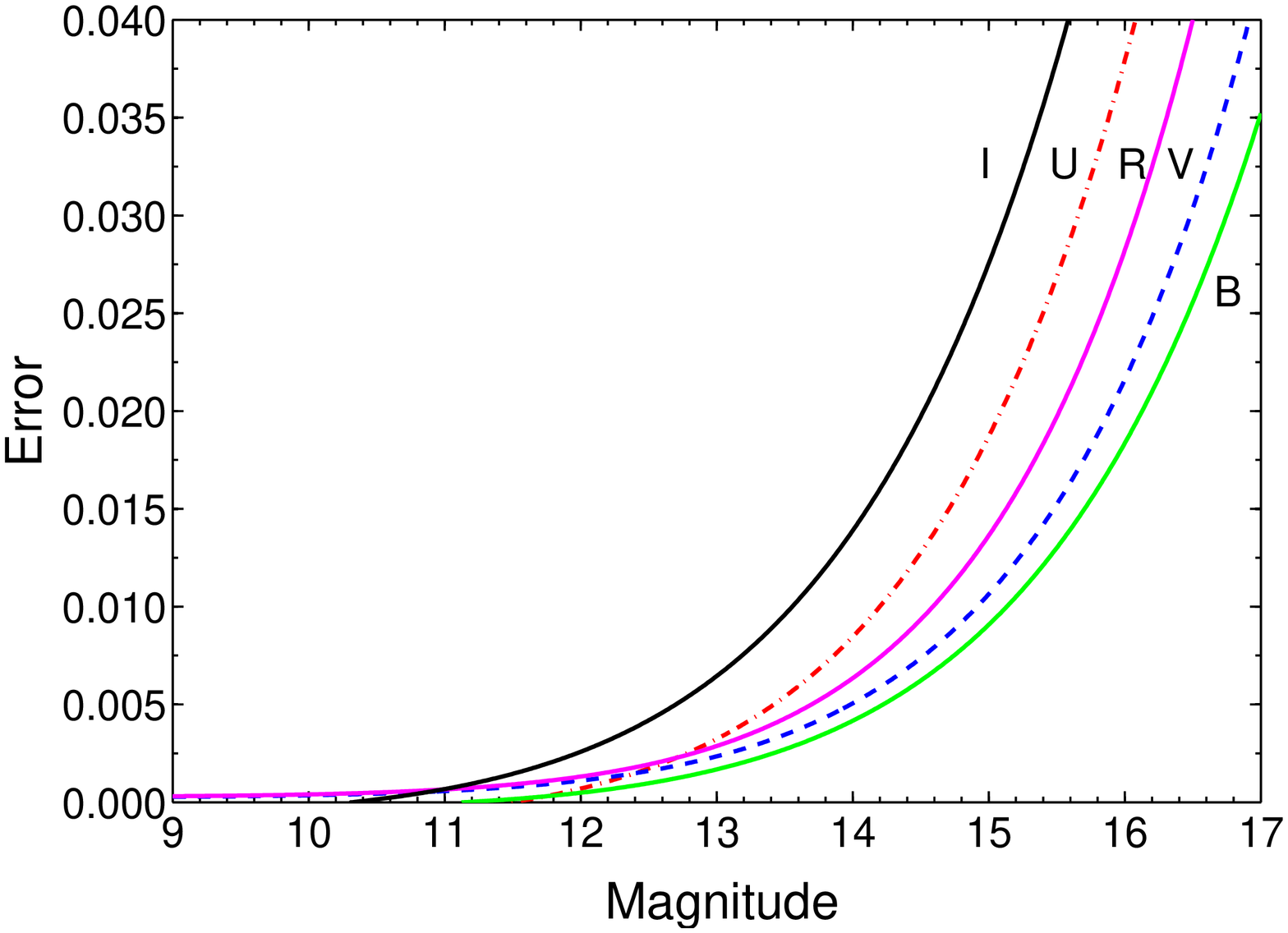}
   \caption{Photometric errors of 73 Landolt standard stars obtained by
   the TNT system. Symbols in left panel show the distribution of the observed
   values in the $UBVRI$ filters while the curves in the right panel
   represent the best fit to the corresponding data points.}
   \label{Fig:prec}
\end{figure}

We further estimated the photometric precision of the TNT system. The errors of 1684
data points for the observations of 73 Landolt standard stars are shown in
Fig. \ref{Fig:prec}. It is clearly seen that the photometric precision
is $\leq$ 0.01 mag for sources brighter than 15.0 mag, with an exposure time of
300-600~s in $U$, 60-120~s in $B$, 40-90~s in $V$, 20-60~s in $R$, and 20-40~s in $I$,
respectively. This is similar to the precision reached by the 85-cm telescope for
a similar exposure time and SNR (Zhou et al. \cite{Zhou09}).

\section{SUMMARY}\label{sect:sum}
In this article, we evaluate performance of the VeryArray:1300B CCD photometric
system mounted on the Tsinghua-NAOC 0.8-m telescope at Xinglong Observatory of NAOC. The evaluation
results are summarized as follows:

(1) Typical CCD parameters such as the bias, gain, readout noise, and the dark current
are derived for VeryArray:1300B. These parameters, especially the bias and the readout noise, are
related to the readout modes. Compared with the fast readout mode, the slow mode produces an
apparently lower bias level and readout noise. Because of a very low working temperature, the
dark current of the CCD detector is very low and can be ignored in the image reduction.

(2) Based on the observations of the Landolt's standard stars on a dozen of
photometric nights, we derived the transformation coefficients
between the instrumental $ubvri$ magnitudes and the standard $UBVRI$ magnitudes:
(i) the color terms used to normalize the photometry are relatively small
 for the $ubvri$ filters mounted on TNT, suggesting that the response curves are similar to
 those of the standard Johnson/Cousins (Bessel) system; (ii) the atmospheric extinction coefficients
 in $UBVRI$ bands is robustly determined for the site of Xinglong with our extensive calibration
 data taken on the photometric nights.

(3) The limiting magnitudes are also obtained for TNT. With an exposure time of 300~s,
it can detect a point source with B$\sim$19.0 mag and V$\sim$18.8 mag for a SNR$\sim$100.

(4) The emission of the sky background at Xinglong Observatory was also examined with our extensive
calibration data, which shows an apparent increase after year 2005, e.g. from a level of V$\sim$21.4
mag in year 2005 to V$\sim$20.2 mag in year 2011. This change definitely brings a negative effect on
the astronomical observations and researches at Xinglong Observatory.

{\it \bf Acknowledgments\ \ }  We thank the anonymous referee for his/her suggestive comments that help 
improve the manuscript. The work here is supported by the National Natural Science Foundation of China 
(NSFC grants 11178003, 11073013, and 10173003) and the National Key Basic Research Science Foundation 
(NKBRSF TG199075402).

\label{lastpage}


\begin{thebibliography}{99}
  \bibitem[1990]{Bessell90} Bessell, M.~S., 1990, \pasp, 102, 1181
  \bibitem[2012]{Fang12} Fang, X. S., Gu, S. H.,  et al., 2012, \raa, 12, 93
  \bibitem[2009]{Fu09} Fu, J. N., Zha, Q., et al., 2009, \pasp, 121, 251
  \bibitem[2000]{Howell00} Howell, S. B., 2000, Handbook of CCD Astronomy (ISBN 0-521-64834-3), Cambridge University Press
  \bibitem[2005]{Kino05} Kinoshita, D., Chen, C.W., Lin, H.C., et al., 2005, \chjaa, 5, 315
  \bibitem[1992]{Landolt92} Landolt, A. U., 1992, \aj, 104, 340
  \bibitem[2009]{Li09} Li, H. L., Yang, Y. G., Su, W., Wang, H. J., Wei,J. Y., 2009, \raa, 9, 1035
  \bibitem[2010]{Liu10} Liu, H., Wang, J., Mao, Y. F., Wei, J. Y., 2010, \apj, 715, 113
  \bibitem[2003]{Liu03} Liu, Y., Zhou, X., Sun, W. X., et al., 2003, \pasp, 115, 2003
  \bibitem[1998]{Shi98} Shi, H. M., Qiao, Q. Y., Hu, J. Y. et al., 1998, Acta Astrophysica Sinica, 18, 99
  \bibitem[2002]{Str02} Stritzinger, M., Hamuy, M., et al., 2002, \aj, 124, 2100
  \bibitem[2008]{Wang08} Wang, X. F., Li, W. D., Filippenko, A. V. et al., 2008, \apj, 675, 626
  \bibitem[2009]{Wang09} Wang, X. F., Li, W. D., Filippenko, A. V. et al., 2009, \apj, 697, 380
  \bibitem[2012]{Wang12} Wang, X. F., et al., 2012, in preparations
  \bibitem[2005]{Wu05} Wu, C., Qiu, Y. L., Deng, J. S., et al.,  2005, \aj, 130, 1640
  \bibitem[2006]{Wu06} Wu, C., Qiu, Y. L., Deng, J. S., et al.,  2006, \aap, 453, 895
  \bibitem[2010]{Xin10} Xin, L.P., Zheng, W. K., et al., 2010, \mnras, 401, 2005
  \bibitem[2011]{Xin11} Xin, L.P., Liang, E. W., et al., 2011, \mnras, 410, 27
  \bibitem[2000]{Yan00} Yan, H. J., et al., 2000, \pasp, 112, 691
  \bibitem[2012]{Yan12} Yan, J. Z., Li, H., Liu, Q. Z., 2012, \apj, 744, 37
  \bibitem[2010]{Yang10} Yang, Y. G., Wei, J. Y., Kreiner, J. M., Li, H. L., 2010, \aj, 139, 195
  \bibitem[2011]{Zhai11} Zhai, M., Zheng, W. K.,  Wei, J. Y., 2011, \aap, 531, 90
  \bibitem[2012]{Zhai12} Zhai, M., Wei, J. Y., 2012, \aap, 538, A125
  \bibitem[2010]{Zhang10} Zhang, T. M., Wang, X. F., et al., 2010, \pasp, 122, 1
  \bibitem[2012]{Zhang12} Zhang, T. M., et al., 2012, in preparations
  \bibitem[2009]{Zhou09} Zhou, A. Y., Jiang, X. J., Zhang, Y. P., Wei, J. Y.,  2009, \raa, 9, 349

\end{thebibliography}
\end{document}